\documentclass[10pt,journal,compsoc]{IEEEtran}

\usepackage{graphicx}
\usepackage{color}
\usepackage{cite}
\usepackage{soul}
\usepackage{amsfonts}
\usepackage[utf8]{inputenc}
\usepackage{algpseudocode}
\usepackage{subfigure}
\usepackage{multirow}
\usepackage{fancyhdr}
\usepackage{amsthm}
\usepackage{textcomp}
\usepackage{physics}
\usepackage{mdwmath}
\usepackage{bm}
\usepackage{mathrsfs}
\usepackage{adjustbox}
\usepackage{commath}
\usepackage{algorithm}
\usepackage{float}

\usepackage{esvect}

\begin{document}
\title{A Study of Language and Classifier-independent Feature Analysis for Vocal Emotion Recognition}

\author{Fatemeh Noroozi,
Marina Marjanovic, 
Angelina Njegus,
Sergio Escalera,
Gholamreza Anbarjafari,~\IEEEmembership{Senior~Member,~IEEE}
\IEEEcompsocitemizethanks{\IEEEcompsocthanksitem F. Noroozi was with Institute of Technology, University of Tartu, Estonia.\protect\\
\IEEEcompsocthanksitem M. Marjanovic and A. Njegus are with Faculty of Technical Sciences, Singidunum University, Belgrade 11000, Serbia.\protect\\
\IEEEcompsocthanksitem S. Escalera is with the Computer Vision Center , University of Barcelona, Barcelona and University of Autonoma, Barcelona, Spain.\protect\\
\IEEEcompsocthanksitem G.~Anbarjafari is with the the iCV Research Lab, Institute of Technology, University of Tartu, Tartu, Estonia. He is also with Department of Electrical and Electronic Engineering, Hasan Kalyoncu University, Gaziantep, Turkey.\protect\\
E-mail: shb@icv.tuit.ut.ee
}}

\markboth{ArXiv, November 2018}%
{Language and Classifier Independent Emotion Recognition}

\IEEEtitleabstractindextext{%
\begin{abstract}
Every speech signal carries implicit information about the emotions, which can be extracted by speech processing methods. In this paper, we propose an algorithm for extracting features that are independent from the spoken language and the classification method to have comparatively good recognition performance on different languages independent from the employed classification methods. The proposed algorithm is composed of three stages. In the first stage, we propose a feature ranking method analyzing the state-of-the-art voice quality features. In the second stage, we propose a method for finding the subset of the common features for each language and classifier. In the third stage, we compare our approach with the recognition rate of the state-of-the-art filter methods. We use three databases with different languages, namely, Polish, Serbian and English. Also three different classifiers, namely, nearest neighbour, support vector machine and gradient descent neural network, are employed. It is shown that our method for selecting the most significant language-independent and method-independent features in many cases outperforms state-of-the-art filter methods. 
\end{abstract}

}

\maketitle

\IEEEdisplaynontitleabstractindextext

\IEEEpeerreviewmaketitle

\section{Introduction}
\label{sec:introduction}
\begin{figure*}[t] 
	\centering
	\includegraphics[width=0.9\textwidth]{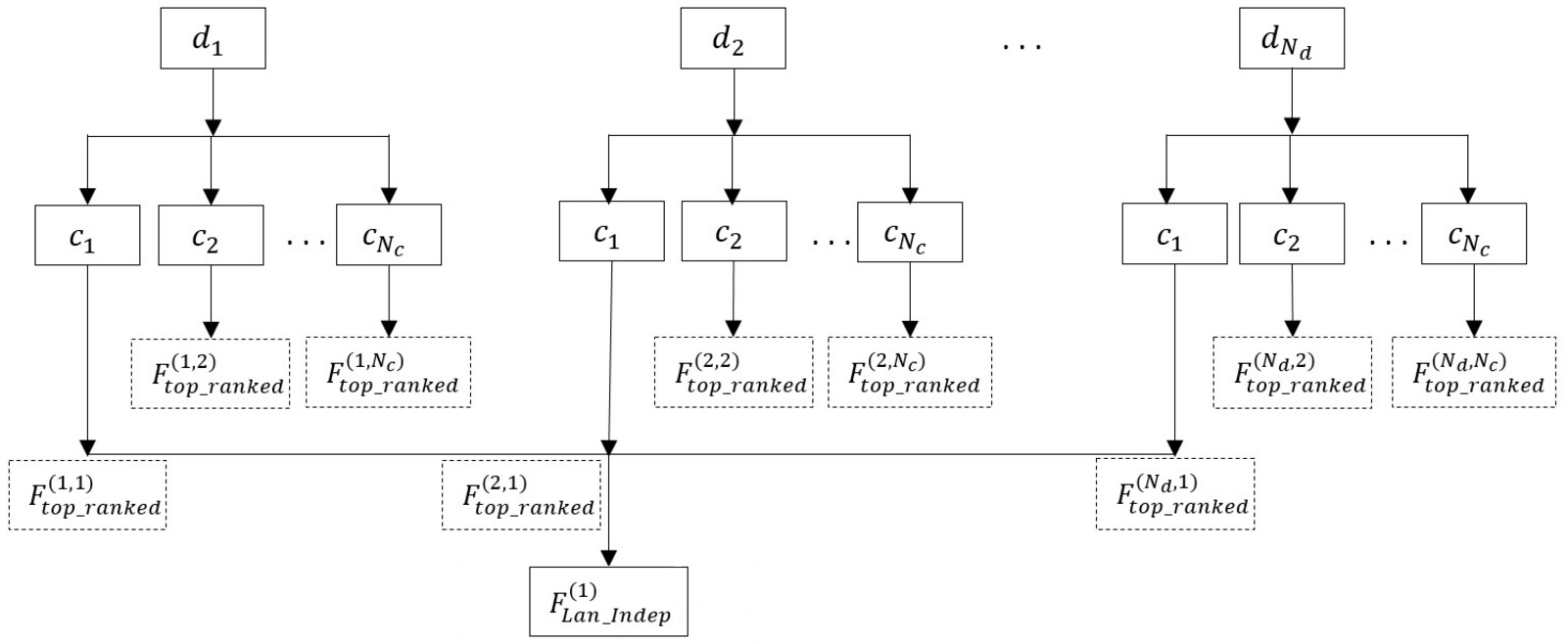}
	\caption{Feature selection strategy flowchart for language-independent analysis.}
	\label{fig4}
\end{figure*}

\begin{figure*}[t] 
	\centering
	\includegraphics[width=0.9\textwidth]{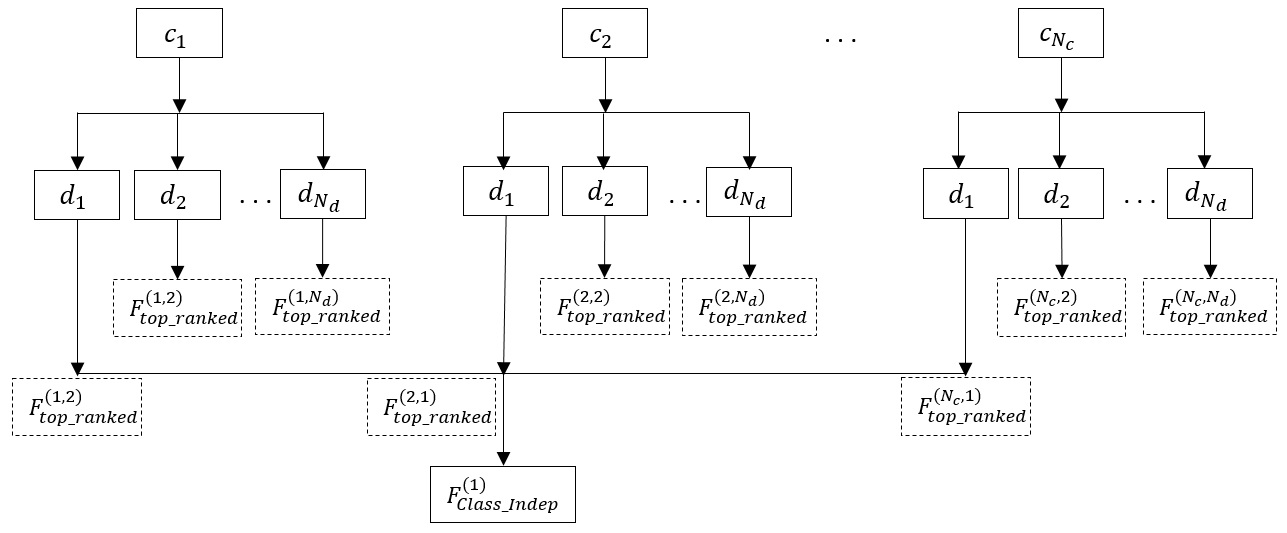}
	\caption{Feature selection strategy flowchart for classifier-independent analysis.}
	\label{tt1}
\end{figure*}

\IEEEPARstart{U}nderstanding the verbal communication of emotion is still a challenging task in the field of human-computer interaction~(HCI)~\cite{helander2014handbook}. Machine learning is a fundamental component of HCI~\cite{rautaray2015vision,demirel2009data,haamer2018changes,gorbova2018integrating,kulkarni2018automatic}. In order to make HCI more realistic, the computer or any intelligent system needs to be able to recognize the emotional state of the human who is interacting with such a system. Researchers have made efforts to find ways for making automatic emotion recognition systems~\cite{mencattini2014speech,corneanu2018survey}. \\
Besides emotion signals, such as facial expressions, gesture recognition, eye contact or other body languages, identification of the emotion from only vocal expressions is harder to recognize. The main reason is that speech consists of two simultaneous components, such as linguistic (what is said) and paralinguistic features (how it is said). 
A vocal emotion recognition can be involved in many applications like e-learning where the state of the learner could be detected in order to adjust the material or the presentation style of an online tutor \cite{el2011survey,wiggins2014relationship}. Today, vocal emotion recognition reaches broader commercial interests, such as employee mood identification \cite{tsai2001determinants,gorbova2017automated}, the game industry \cite{vogt2008emovoice}, or call centres \cite{yacoub2003recognition}, where the analysis of users’ emotional states could say much about clients' satisfaction \cite{dai2015emotion}. \\
A vocal emotion recognition system consists of speech signal input, signal preprocessing, spectral analysis, feature extraction, emotion classification, and pattern recognition \cite{dai2015emotion}. Humans produce speech signals, which comprise the basis of acoustics interaction. Speech signals convey the lexical and paralinguistic information which carry out the contents and emotions. The lexical information includes the conceptual structure that is related to the language. In contrast, paralinguistic features are not dependent on the semantic structure of the language~\cite{schuller2013paralinguistics,chen2012speech}. They can be extracted by using signal processing techniques to infer the human emotions based on speech signals, i.e. for vocal emotion recognition~\cite{noroozivocal,zarate2015multiple,kaminska2017efficiency}. Since numerous paralinguistic features exist, selecting a strong and suitable set of features is a challenging task. A large number of extracted features usually increases the computational complexity and the classification error. Thus it is important to eliminate irrelevant and correlated features.\\
The features should be able to efficiently distinguish between different emotions. Additionally, they should be independent from the lexical or language-related contents of the words, since language-independent features are more robust and reliable when dealing with different languages. Furthermore, when using language-independent features, the automatic vocal emotion recognition system can be applied to databases with different languages. Otherwise, if language-dependent features are used, the system has to be changed every time in order to tailor it to the specifications of the language of the database. The features that have been used in the recently published research are pitch, intensity~\cite{alonso2015new},  formants~\cite{tamuri2015expression}, mel-frequency cepstral coefficients (MFCCs)~\cite{ayoub2015analysis} and filter bank energies~(FBE)\cite{pan2012speech}.\\
The majority of previous studies have exploited different types of features and classifiers. Recently, researchers have aimed at improving the performance rate and the computational cost of the automatic vocal emotion recognition system by selecting uncorrelated features that could reduce the possibility of misclassification~\cite{anagnostopoulos2015features}. Schull et al. use the dynamic base contour and filters such as wrapper based search, to create a feature set from features such as formants, MFCCs, intensity, and pitch \cite{schuller2006evolutionary}, and to construct feature vectors. They have used classifiers such as support vector machine (SVM)~\cite{tan2010learning},  k-nearest neighbor (KNN)~\cite{pao2012study} and Naive Bayes~\cite{tomavsev2014hubness} to test the result of classification with selected features. They have achieved the best performance by using MFCCs with the SVM classifier. Wagner et al. \cite{wagner2005physiological} used the principal component analysis~(PCA) for feature selection, where the feature selection problem is investigated from a physiological point of view. They used well-known classifiers such as KNN and Multi-layer perceptrons Neural Networks~(MLP) to compare the recognition performance rate before and after feature selection. Kostoulas et al. \cite{kostoulas2010enhancing} and Anagnostopoulos et al. \cite{anagnostopoulos2009sound} have used a correlation based subset evaluator to determine the optimal subset of the features. Although these features improve the recognition rate, they are not independent from the language.\\
Languages are different in terms of grammatical and morphological properties. Actually the culture of the region that the language belongs to also affects the tone and this is called the "pragmatic" aspect. According to linguistics research, different dialects  also affect the acoustics factors~\cite{laukka2014evidence}. In addition, this influences the overall trend and strength by which the emotional expression shows its impact on the quantitative measures, i.e. the acoustic features. These characteristics of the voice convey the meanings of the spoken words and phrases. They are essential for expressing the feelings, intentions and emotions using particular patterns in the variations of paralinguistic features. This can affect the properties of every language~\cite{el2011survey}. The previous works show the undeniable effects of paralinguistic features in introducing emotions into speech and hierarchical communication~\cite{zarate2015multiple}.\\
In this paper, the interaction between humans and computers is investigated only based on acoustic features, such as pitch, intensity, formants, MFCCs and FBEs, together with other features such as autocorrelation~\cite{markel2013linear},  minimum \cite{loizou2013speech}, maximum \cite{loizou2013speech}, variance~\cite{li2014overview}, standard deviation \cite{schuller2003hidden}, percentiles~\cite{kwon2003emotion}, zero crossing rate (ZCR) \cite{madisetti2009digital} and  ZCR density \cite{tanyer2000voice}. In total, 84 features have been extracted by using signal processing techniques. The main contribution of the proposed method is that the features used by the proposed method are independent from the language of the database and the type of classifier. The classifiers that are used in this paper, in order to determine the features which are independent from the choice of classifier, are multi-class SVM (M-SVM), KNN, and deep learning Neural Networks \cite{deng2014deep}.\\
In order to verify the independence of the features from the language content, speech signals from three different languages, namely, Polish, English and Serbian, are utilized. The mentioned languages are categorized as Indo-European~\cite{sapir2014language}. The Polish and the Serbian languages are from the Slavic category, and English is a Germanic language \cite{marinova2013association}. The Polish emotional speech database \cite{staroniewicz2008polish}, the Surrey Audio-Visual Expressed Emotion (SAVEE) \cite{jackson2014surrey} dataset and the Serbian Emotional dataset \cite{jovicic2004serbian} are used.\\
The closest research work to this paper is a work on language-independent feature extraction proposed in~\cite{shaukat2010exploring}. They use four different datasets and $87$ features and two classifiers. Although some of the features that we used are the same as their features, such as formants, pitch, intensity and MFCCs, we have used FBEs as a feature of the speech signal, and it gave good results. Also in this work a systematic justification on the choice of the features is provided. This work also contributes by using deep learning Neural Networks to investigate the recognition rate by using different training sets.\\
Another contribution of this work compared to previous works is that in this work we are also focusing on finding classifier-independent features. Different performance rates are obtained by different classification algorithms on the same constructed databases by using the same combinations of features.\\
We apply different classification methods to the proposed combination of language-independent features, in order to compare their performances, based on the recognition rates. We also use a Convolutional Neural Network (CNN)~\cite{bottou2010large,le2015tutorial}, in order to evaluate the performance of the features extracted by using deep learning methods with the acoustic features, in terms of vocal emotion recognition efficiency. However, since CNNs usually expect images as their inputs, we extract the related spectrograms from the speech signals, which are represented as images, and can be used as inputs to the CNN.
The remainder of the paper is structured as follows. The proposed method is described in Section \ref{s2}. Then the experimental results are presented and discussed in Section \ref{s3}. Finally, the paper is concluded in Section \ref{s4}. The general block diagrams of procedures in this paper are shown in Figures \ref{fig4} and \ref{tt1} and the description of the details is produced in Section 2.


\section{Proposed Method}\label{s2}
In this section, we present the proposed approach for selecting the features which result in the development of language-independent vocal emotion recognition. For this purpose, feature extraction is first explained. Then the feature selection model is introduced, and finally a description of adopted classifiers is given.
\subsection{Feature Extraction}
Making a reliable dataset for the recognition problem is the first step. Inefficient and insufficient choices of the features can cause overlaps and misclassification~\cite{anagnostopoulos2015features}. As mentioned before, in the paralinguistic approach, by its nature, the acoustic signals could be used to extract numerous features by applying signal processing methods. In this research, 84 features have been extracted.\\
The paralinguistic elements of the voice such as loudness, speed and other elements, usually change between different languages. The paralinguistic features utilized by the proposed method are listed as follows with brief definitions: \begin{itemize}
\item Pitch is known as a prosodic feature which can be characterized by the fundamental frequency~\cite{hess2012pitch,lee2015audio}. It is the lowest frequency component, and contains speaker-specific information. Pitch is shown as ${\rho_0}(s)$. It has been used in many studies on vocal emotion recognition, such as \cite{globerson2013psychoacoustic,pan2012speech}. 
\item Intensity refers to the loudness of the speech signal $s$. It is measured at the position of the syllable peak, which is most commonly a vowel~\cite{harrington2010phonetic}. Examples of articles that utilized intensity as an important feature for vocal emotion recognition are \cite{laukka2014evidence,chronaki2015development}.
\item The length of the vocal tract can be defined as a useful feature. Longer vocal tracts produce lower resonating frequencies which are known as formant frequencies. In \cite{siegert2014investigation}, the length of the vocal tract was used for recognizing emotions based on speech.
\item Standard deviation of the speech signal $s$, ${std}$, is also used as one of the acoustic features \cite{allgood2015developmental}.
\item MFCCs can be calculated according to the instructions given in \cite{Xphdthesis}. 
MFCCs are well-known vocal features that have been used for emotion recognition in articles such as \cite{dai2015emotion,pan2012speech}.
\item Another function that is used is the zero-crossing rate (ZCR). It counts the variations of the sign of the data. ZCR is an important feature that has been utilized for vocal emotion recognition in \cite{wang2015speech,origlia2014continuous}. The formulation provided in \cite{gouyon2000use} can be used for calculating ZCR.
\item According to~\cite{fant1960acoustic}, formants are the spectral peaks of the sound spectrum of the voice. They are useful for distinguishing the elements of the speech (i.e. the vowels in the sound). They have been used for vocal emotion recognition in papers such as \cite{laukka2014evidence,dai2015emotion}.\\ 
The formant with the lowest frequency is named $f_{1}$, the second $f_{2}$, and the third  $f_{3}$. 
\item The last features category used in this paper is extracted by using the filter banks. In order to filter the speech file,  FBEs, a method from ~\cite{skowronski2003improving}, is applied. The FBEs have been used in studies such as \cite{joshi2013recognition} in order to perform vocal emotion recognition. 
\end{itemize}
For extracting the features, we use signal processing techniques implemented in MATLAB and PRAAT software \cite{boersma2013praat}. It should be noted that silence might also be meaningful and stand for a certain emotion, such as fear. If silence exists in the signal, it appears in the form of continuous zero intensity within the corresponding interval.
\subsection{Classification}
In order to perform classification, we adopt that notation used for the training set, $\{x_i, y_i\}$, where $x_i\in R^{Nf}$ is a vector of extracted features, $i =\{ 1,...,N\}$ (where $N$ represents the number of the samples) and vector ${y_i}$ represents its associated class.\\
In this paper, we use four classification methods. Due to their good performance, we applied two deep learning neural networks, the stochastic gradient descent algorithm and CNN~\cite{bottou2010large,le2015tutorial}.
As a second classifier, as one of the most famous algorithms for the prediction of the class of new samples, we applied the nearest--neighbor rule \cite{weinberger2005distance}. 
As a third classifier, the M-SVM is used, which includes multiple binary SVMs~\cite{cortes1995support}.
\subsubsection{Independent Feature Selection Strategy}
In order to select language-independent features, first we observe three different datasets and three different classifiers. However, this strategy is flexible and can be easily expanded on $N_d$ datasets and $N_c$ classifiers.
Let's consider that $D=\big\{d_1,d_2,...,d_{N_d}\big\}$ represents the set of all datasets, $C=\big\{c_1,c_2,...,c_{N_c}\big\}$ is the set of all classifiers, and $F=\big\{x_1,x_2,...,x_{N_f}\big\}$ represents the set of all extracted features.\\ 
\textit{a) language-independent Features Selection Strategy}

In order to find a subset of language-independent features, our objective is to form a subset ${F^{(i,j)}_{top-ranked}}$ of $m$ top-ranked features from the set $F$ for the dataset $d_i$ and the classifier $c_j$. This is achieved in the following ways:
\begin{itemize}
    \item Our approach by testing the dataset $d_i$ with classifier $c_j$ just taking in account of each feature separately and comparing their recognition rates;
    \item In \cite{njegus2016}, 5 widely used filter methods are summarized and compared for the Serbian corpora:
    \begin{enumerate}
        \item Gain Ratio (GR) that evaluates the weight of a feature by measuring the gate ratio with respect to the class;
        \item Information Gain (IG) evaluates the weight of a feature by measuring the information gain with respect to the class;
        \item Correlation-based Feature Selection (CFS) evaluates the weight of a feature by measuring the correlation between it and the class;
        \item ReliefF (RF) evaluates the feature weight by repeatedly sampling an instance and considering the value of the given feature for the nearest instance of the same and different classes;
        \item Symmetrical Uncertainty (SU) that evaluates the feature weight by measuring the symmetrical uncertainty with respect to the class.
    \end{enumerate}
\end{itemize}
After the features ranking, the subset ${F^{(i,j)}_{top-ranked}}$ can be represented as:
\begin{equation}
\label{tt}
{F^{(i,j)}_{top-ranked}}=\big\{{x^{(i,j)}_{1}},{x^{(i,j)}_{2}},...,{x^{(i,j)}_{m}}\big\}
\end{equation}
 ${x^{(i,j)}_{m}}$ is the $m^{th}$ feature of the dataset $d_i$ where classifier $c_j$ is applied. Observing the intersection of subsets  ${F^{(1,j)}_{top-ranked}}$, ${F^{(2,j)}_{top-ranked}}$... and ${F^{({N_d},j)}_{top-ranked}}$, represented as:
 \begin{equation}
{F^{(j)}_{lan-indep}}=\bigcap_{i=1}^{N_d}{F^{(i,j)}_{top-ranked}}
\end{equation}
we are selecting the set of common features for the classifier $c_j$. Those features are considered as \textquotedblleft language-independent\textquotedblright.
At the end, the  subset of language-independent features is tested using all classifiers.
Figure \ref{fig4} illustrates the language-independent feature selection for the first classifier.\\
\textit{b) Classifier-independent Features Selection Strategy}

Similar to language feature selection, in order to find a subset of classifier-independent features, our objective is to form a subset $F^{(i,j)}_{top-ranked}$ of $m$ top-ranked features from the set $F$ for the classifier $c_i$ and the dataset $d_j$. The same as in the language-independent feature selection strategy, this is achieved by testing the dataset $d_j$ with classifier $c_i$ just taking into account each feature separately and comparing their recognition rates. \\
In the case of classifier-independent feature selection, using filter methods is not possible for feature ranking, since those methods are independent from the classifier.
Therefore, the subset $F^{(i,j)}_{top-ranked}$ can be represented as in equation \ref{tt}, where ${x_m}^{(i,j)}$ is the $m^{th}$ feature of the dataset $d_j$ where classifier $c_i$ is applied. Observing the intersection of subsets $F^{(1,j)}_{top-ranked}$, $F^{(2,j)}_{top-ranked}$, ... and $F^{(N_c,j)}_{top-ranked}$ represented as:
\begin{equation}
{F^{(j)}_{class-indep}}=\bigcap_{i=1}^{N_c}{F^{(i,j)}_{top-ranked}}.
\end{equation}
we are selecting the set of common features for the dataset $d_j$. Those features are \textquotedblleft classifier-independent\textquotedblright. At the end, the  subset of classifier-independent features is tested using all classifiers.
Figure~\ref{tt1} illustrates the classifier-independent feature selection for the first dataset.\\
Finally, to obtain both a language- and classifier-independent subset of features, we need to find a language-independent features subset for all classifiers $F_{lan-indep}$, i.e:
\begin{equation}
{F_{lan-indep}}=\bigcap_{j=1}^{N_c}{F^{(j)}_{lan-indep}}
\end{equation}
which is the same subset as a classifier-independent features subset for all languages, i.e:
\begin{equation}
{F_{class-indep}}=\bigcap_{j=1}^{N_d}{F^{(j)}_{class-indep}}.
\end{equation} 
At the end, the  subset of independent features is tested using all classifiers.
Additionally, in the result section, we will show the performance of each classifier considering  {\textquotedblleft special features\textquotedblright}   for classifier $j$  that are made as a union of top $p$ ($p$$<$$m$) ranked features for each dataset $i$, i.e:
\begin{equation}
{F_{spec-feat}}=\bigcup_{j=1}^{N_d}{F^{(j)}_{top-ranked}}
\end{equation}

\subsubsection{Most Affective Features Selection Strategy}  
In order to evaluate the effect of every feature in recognizing each of the emotions, every time, a certain classifier and a particular feature are considered, and the overall recognition rate is calculated for each of the emotions separately. For every emotion, the database is split into two sets, where one of the sets consists of all the samples that represent the considered emotion, and the other one consists of the rest of the samples. The leave-one-out method is used for cross validation. After doing so on all the emotions, a recognition rate is available for every emotion, i.e. seven values in total, which are helpful information for assessing the level of suitability of the particular feature and classifier under study for recognizing each of the emotions. The proposed method has been tested by using the AdaBoost classifier~\cite{freund1999short} and the Serbian database.

\subsubsection{Classification by Using CNN}

\begin{figure*}[t] 
	\centering
	\includegraphics[width=0.9\textwidth]{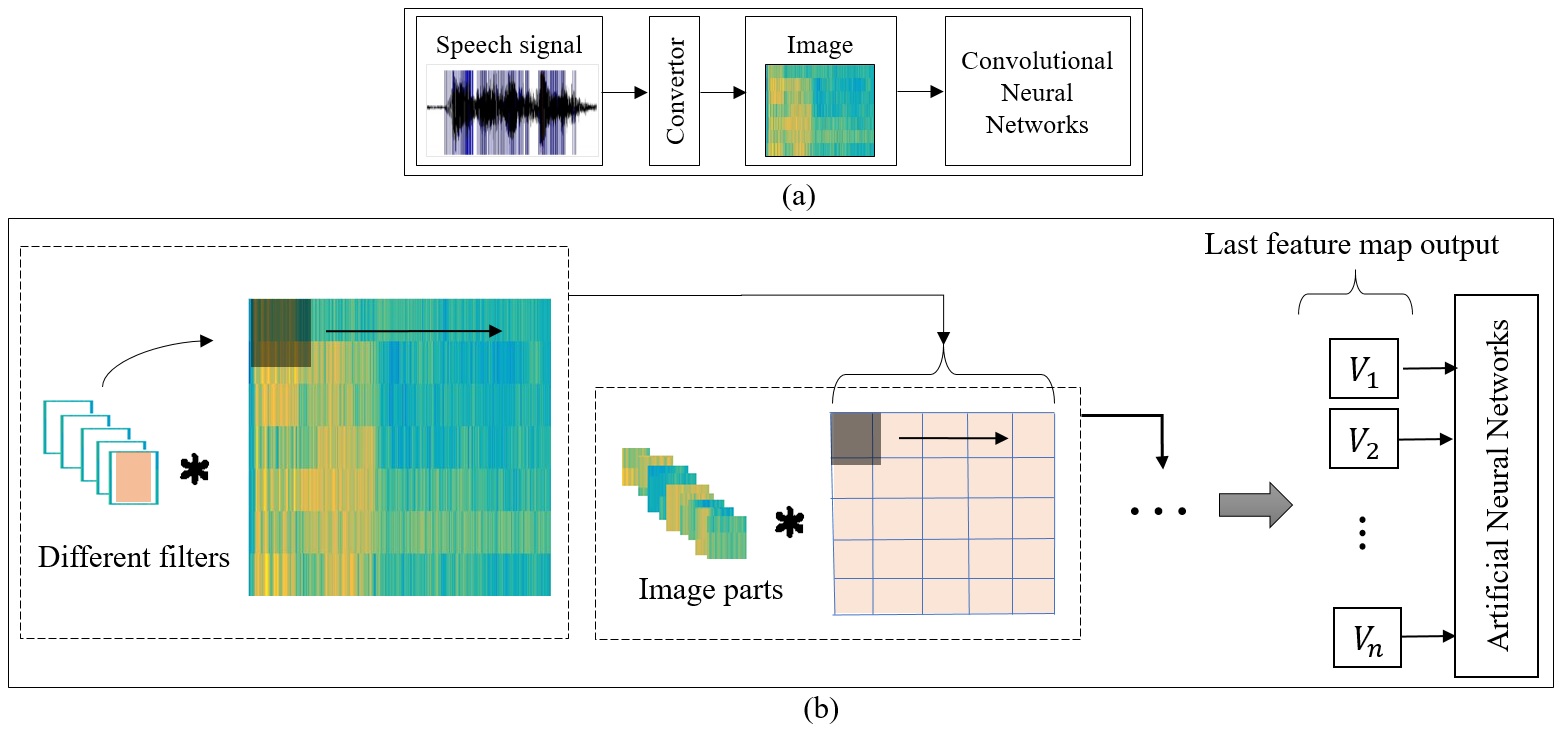}
	\caption{The process of using a CNN for speech-based emotion recognition: (a) shows the process of making the spectrogram images based on the speech signals, and (b) shows the structure of the CNN classifier, which takes the spectrogram images as its inputs.}
	\label{figlast1}
\end{figure*}

CNNs are strong, state-of-the-art deep learning tools for pattern recognition tasks, including classification of signals representing different emotional states. This is because of their well configured structures, consisting of multiple layers of neural networks. It enables them to determine the most distinctive features based on enormous collections of data~\cite{bhandare2016applications}. In this section, we aim to compare the distinctiveness of the proposed set of language-independent, paralinguistic acoustic features with CNN-based features that are independent from the contents of the speech signals. Therefore, we utilize a CNN for performing classification on the same databases, i.e. Polish, SAVEE and Serbian. However, as CNNs usually take images as their inputs~\cite{schmidhuber2015deep}, we extract spectrograms from the speech signals, which can be converted to images. They provide proper representations of the speech signals, and can be used as inputs to the CNN.\\
The CNN considers each image as an $n\cross{n}$ matrix, and uses the convolution operator in order to implement a filter vector. The output of the first convolution will be a new image, which will be passed through another convolution by a new filter. This procedure will continue until the most suitable feature vector elements $\{V_1,V_2,...,V_n\}$ are found. Next, by using a neural network, the probability of each emotion class is calculated. The procedure of using a CNN for speech-based emotion recognition is shown in Fig.~\ref{figlast1}.    

\section{Experimental Results and Discussion} \label{s3}

\subsection{Features Explanation}
In this section, for experimental purposes, ${N_f}=84$ state-of-the-art features have been used. Those features are commonly used in vocal based emotion recognition due to the important characteristics that they introduce in vocal based emotion recognition. According to \cite{nadeu2001time}, frequency-filtered FBEs are robust features that preserve the information carried by the frequencies. They conclude that good time and frequency filters can increase the recognition performance rate. FBEs are used in many papers such as \cite{wolf2014channel,esmaili2016automatic} to achieve high recognition performance.\\
MFCC is used in \cite{delic2012discrimination} as a suitable spectral feature. In \cite{rao2013robust}, it is mentioned that MFCC correlates with the information about the vocal tract. It has been also mentioned that MFCC considers the non-linear auditory perceptual system of the speaker, which is helpful for automatic vocal recognition.\\
Formants are also discussed in \cite{rao2013robust} as important features for speech-based emotion recognition. It has been shown that the sharpness of formants has distinctive properties which can be used for emotion recognition. In \cite{ingram2013formant}, it is shown that formants contain information about the static and dynamic properties of the speech, which show many aspects, such as vocal tract morphology, articulatory setting, dialect and speaking style, which are useful for emotion recognition.\\
In \cite{schuller2011recognising}, the significance of pitch and intensity have been shown. Schuller et al. have concluded that the emotion recognition systems that use pitch and intensity are usually successful because of the supra-segmental nature of emotional speech signals. According to \cite{delic2012discrimination}, the emotional state of the speaker changes the tensions of the vocal cord and the sub-glottal air pressure, which affect pitch. Thus it can be used for emotion recognition.\\
According to \cite{fernandez2011recognizing,schuller2011recognising}, standard deviation, minimum, maximum, percentiles, autocorrelation, variation and mean are important statistical features. Some of these features are also used for extracting other features. For example, autocorrelation has been used in \cite{shimamura2001weighted} for extracting the pitch.\\
In \cite{lee2014detecting}, the harmonic-to-noise ratio has been discussed. Despite the static harmonic-to-noise ratio of sustained vowels, they consider the dynamic movements of the articulatory organs. They have reported that by using the harmonic-to-noise ratio contour modeling, the classification error will be reduced because at limited frequency bands, they can be used to distinguish the pathologic voices \cite{moran2006telephony,gelzinis2008automated}.\\
In \cite{eyben2013recent}, ZCR has been used for multimedia feature extraction. They have different categories of features. They mention that in the category of waveform features, ZCR is one of the most important features. Also in \cite{jalil2013short}, ZCR is used as a feature for distinguishing voiced and unvoiced parts of speech signals. Also in \cite{shete2014zero} they have used ZCR for distinguishing and separating these parts.
Those features, with the label that we use in this paper, are listed in Table~\ref{T1}.

\subsection{Features Ranking Strategy}
\begin{figure*}[t] 
	\centering
	\includegraphics[width=0.9\textwidth]{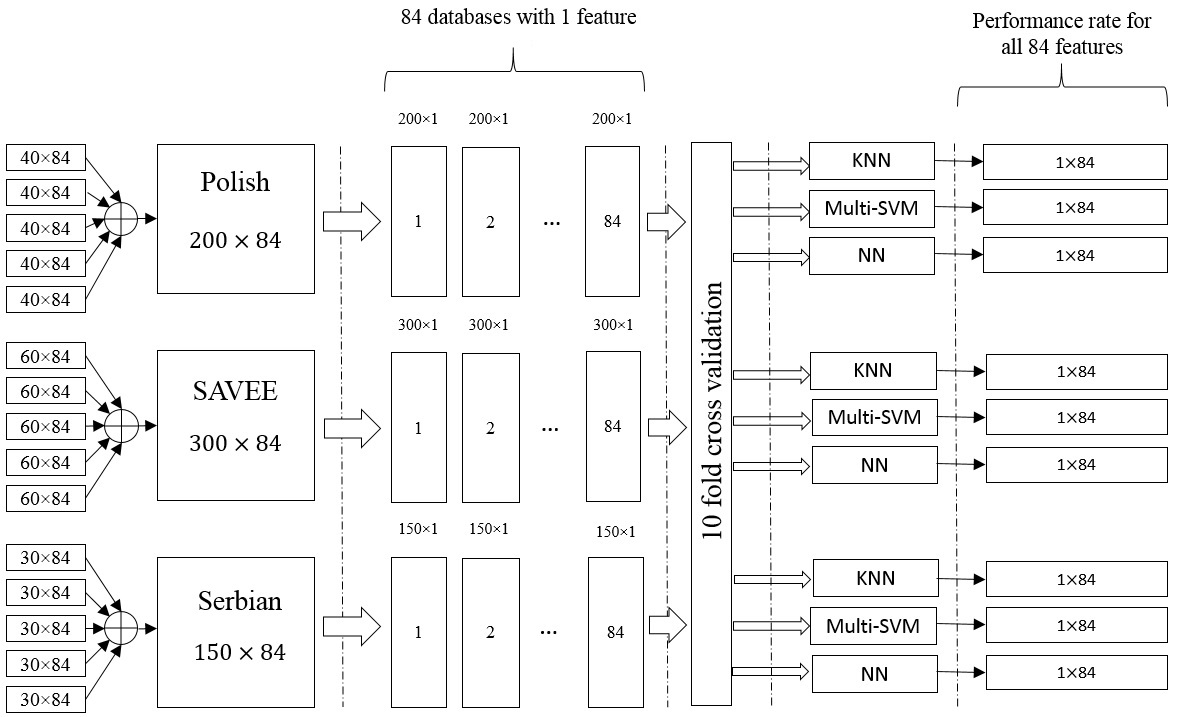}
	\caption{Dataset construction and the classification strategy.}
	\label{fig51}
\end{figure*}

In order to find a subset of classifier-independent features, we form a subset of $m=22$ top-ranked features from each data set for all used classifiers.\\
This is achieved in two ways:\\
a) First, by using our approach, i.e. by testing each dataset with each classifier and comparing their individual performances in terms of recognition rates. In order to calculate the recognition rate for each feature separately, ${N_c}=3$ different classification methods, namely, KNN, M-SVM and Neural Networks with gradient descent, are used. KNN is chosen because it is a simple and efficient classifier, M-SVM is selected due to its speed and reasonable performance, and Neural Networks with gradient descent is picked due to its good performance.\\
b) Second, by using filter methods such as GR, IG, RF and SU only for language-independent feature selection, since in filter methods the subset selection procedure is independent from the classifier and is not applicable on classifier-independent feature selection.\\
In our experiment, we use $N_d=3$ datasets, namely, Polish, SAVEE (English) and Serbian. Five emotional states (anger, fear, neutral, happiness and sadness) are considered, and they are balanced within all corpora.  Every emotion is assigned to $40$ sample vectors in the Polish dataset, $60$ samples in the SAVEE database and $30$ samples in Serbian database. Due to the number of features that we use in experiments, every sample vector has $84$ elements. Therefore, each element represents a feature that we extracted from all audio files, as shown in Fig.~\ref{fig51}.\\
In the next step, each dataset is divided into $84$ parts corresponding to each feature separately. Every new dataset matrix has $(40\times{5}=200)$, $(60\times{5}=300)$ or $(30\times{5}=150)$ rows for Polish, SAVEE or Serbian corpora, respectively. This results in $84$ column vectors with $200$ elements for the Polish dataset, $300$ elements for the SAVEE database and $200$ elements for the Serbian dataset. Therefore, the output of the second step is $84$ datasets which will be input for the three classifiers. We use 10-folds cross validation. Then the performance of every feature for all used classifiers is  individually evaluated.\\ 
Ranked features and the performance rates have been calculated for all possible combinations of languages and classifiers used in this work. 
The results of classification on the Polish database by KNN, M-SVM and Neural Network are illustrated  in the appendix A respectively in Table~\ref{T22}, Table~\ref{T23} and Table~\ref{T24}. These tables show the performance of every feature in the classification process. 
Table~\ref{T25}, Table~\ref{T26} and Table~\ref{T27} in appendix A represent the performance rates on the SAVEE dataset,  and in Table~\ref{T28}, Table~\ref{T29} and Table~\ref{T30},  the performance of features by KNN, M-SVM and Neural Network on the Serbian dataset is shown. 
\subsection{Language-independent Features Selection}
Feature ranking obtained in the previous subsection conducts the feature selection that is performed in the second stage of the proposed algorithm. The ranking of the features is performed in two ways: according to their individual performance and using mentioned filter methods such as GR, IG, RF and SU. The feature selection strategy follows 
the flowcharts shown in Figs.~\ref{fig4} and~\ref{tt1}. Any language dataset can be a case-study of these algorithms. Three sets are obtained through the first step for every language dataset, where every set shows the label of the feature with its performance with one of the mentioned classifiers.
Using individual ranking, features performances for each language and each classifier are shown in the appendices~\ref{AppA} and~\ref{AppB}. For each language, we have three tables for KNN, M-SVM and Neural Networks classifiers, where every table contains the label of the features. According to the ranking from Table~\ref{T22} to Table~\ref{T30}, we present 3 categories of features shown in Table~\ref{T6}, Table~\ref{T7} and Table~\ref{T8} as follows:
\begin{enumerate}
    \item Top selected $22$ features;
    \item According to the flowchart shown in Fig.~\ref{fig4},  from the top selected $22$ features {\textquotedblleft Common features\textquotedblright} is formed;
    \item {\textquotedblleft Special features\textquotedblright} calculated from equation number 22, selecting $p$=10.
\end{enumerate}
In order to compare individual ranking with state-of-the-art filter methods,  features selected using different filter methods, are shown in appendix B from Table~\ref{T2} to Table~\ref{T5}. They are shown for each language, and those methods are independent from the classifier.
In the same appendix according to the Tables, from ~\ref{T2} to ~\ref{T5}, in Table~\ref{T12}, selected language-independent features for English, Polish and Serbian languages, using state-of-the-art filter methods, are shown. On the other hand, in appendix A, Table~\ref{T13}, selected language-independent features for English, Polish and Serbian languages are presented using individual features ranking. We can see that both approaches have in common the following features: Mean of $FBE_{13}$, $MFCC_{1}$, Mean of $MFCC_{1}$, Intensity, $FBE_{3}$, $FBE_{9}$, $FBE_{11}$, Mean of $FBE_{12}$. However, we can always select the optimum features subset according to their performance shown in the next subsection.

\subsection{Classifier-independent Features Selection}
A similar algorithm to the language-independent one is developed for a selection of classifier-independent features. The features that are independent from the classification method have been selected. In this case, the language is kept fixed, and then 22 features with the best performances in classification have been selected. This procedure has been performed on KNN, M-SVM and Neural Networks.\\
In the second step, the common features that are repeated in all three categories are selected. These selected features are independent from the classifier. Then the whole process is repeated for all the languages. According to the ranking from Table~\ref{T22} to Table~\ref{T30} in appendix A, we present three categories of features:
\begin{enumerate}
    \item Top selected $22$ features;
    \item According to the flowchart shown in Figure~\ref{tt1}, {\textquotedblleft Common features\textquotedblright} category from top selected $22$ features is obtained;
    \item {\textquotedblleft Special features\textquotedblright} is calculated from equation number 22, selecting $p$=10.
\end{enumerate}
All three categories of features are presented from Table~\ref{T9} to Table~\ref{T11} in appendix B.\\
In the case of classifier-independent feature selection, we didn't compare individual ranking with state-of-the-art filter methods, since filter methods are independent from the classifier. Finally, in Table~\ref{T14}, selected classifier-independent features for English, Polish and Serbian languages are presented. According to the results shown in Table~\ref{T12}, Table~\ref{T13} and Table~\ref{T14}, the number of features that are independent from the languages is fewer than the features that are independent from the classification method. This shows that the changes of language have stronger effects on the performance of vocal based emotion recognition systems than the changes of classification method.

\begin{table*}[ht]
	\centering
	\caption{All selected language-independent features using state-of-the-art filter methods}
	\label{T12}
	\resizebox{0.9\textwidth}{!}{
	\begin{tabular}{c|l|l|l|l|}
\cline{2-5}
\textbf{Independent features}              & \multicolumn{1}{c|}{\textbf{GR}}                                       & \multicolumn{1}{c|}{\textbf{IG}}                                                                    & \multicolumn{1}{c|}{\textbf{RF}}                                                                                         & \multicolumn{1}{c|}{\textbf{SU}}                  \\ \hline
\multicolumn{1}{|c|}{Language-independent} & \begin{tabular}[c]{@{}l@{}} Mean of $FBE_{11}$\\  $FBE_{9}$\\  $FBE_{10}$\\ Mean of  $FBE_{13}$\end{tabular} & \begin{tabular}[c]{@{}l@{}}Std\\ $FBE_{2}$\\ Mean of  $FBE_{13}$\\ $FBE_{11}$\\ Mean of $FBE_{12}$\\  $FBE_{3}$\\  $FBE_{10}$\end{tabular} & \begin{tabular}[c]{@{}l@{}}Max\\ Intensity\\  $MFCC_{1}$\\ Std\\ Mean of $MFCC_{1}$\\Mean of $FBE_{13}$\end{tabular} & \begin{tabular}[c]{@{}l@{}} Mean of  $FBE_{13}$\\ $FBE_{11}$\\ Mean of $FBE_{12}$\\  $FBE_{3}$\\  $FBE_{10}$\\  $FBE_{9}$\end{tabular}\\ \hline
\end{tabular}
}
\end{table*}

\begin{table}[ht]
	\centering
	\caption{All selected language-independent features using our ranking strategy }
	\label{T13}	
	\resizebox{0.48\textwidth}{!}{
	\begin{tabular}{c|l|l|l|}
\cline{2-4}
\textbf{Independent features}              & \multicolumn{1}{c|}{\textbf{KNN}}                                       & \multicolumn{1}{c|}{\textbf{M-SVM}}                                                                    & \multicolumn{1}{c|}{\textbf{Neural Network}}                                                                                                    \\ \hline
\multicolumn{1}{|c|}{Language-independent} & \begin{tabular}[c]{@{}l@{}}$MFCC_{1}$\\ Mean of $FBE_{13}$\end{tabular} & \begin{tabular}[c]{@{}l@{}}Intensity\\ $FBE_{3}$\\ $FBE_{8}$\\ $FBE_{9}$\\ Mean of $FBE_{13}$\end{tabular} & \begin{tabular}[c]{@{}l@{}}Mean of $MFCC_{1}$\\ $FBE_{9}$\\ $FBE_{10}$\\ $FBE_{11}$\\ Mean of $FBE_{12}$\\ Mean of $FBE_{13}$\end{tabular} \\ \hline
\end{tabular}
}
\end{table}

\begin{table}[ht]
	\centering
	\caption{All selected classifier-independent features using our ranking strategy }
	\label{T14}
	\resizebox{0.48\textwidth}{!}{
	\begin{tabular}{c|l|l|l|}
\cline{2-4}
\textbf{Independent features}            & \multicolumn{1}{c|}{\textbf{Polish}}                                                                                                                                                            & \multicolumn{1}{c|}{\textbf{SAVEE}}                                                                                                                                                                            & \multicolumn{1}{c|}{\textbf{Serbian}}                                                                                                                        \\ \hline
\multicolumn{1}{|c|}{Method-independent} & \begin{tabular}[c]{@{}l@{}}Intensity\\ Standard deviation\\ Minimum\\ Variance\\ Maximum\\ $MFCC_{1}$\\ $FBE_{5}$\\ $FBE_{8}$\\ $FBE_{9}$\\ Mean of $FBE_{8}$\\ Mean of $FBE_{13}$\end{tabular} & \begin{tabular}[c]{@{}l@{}}Standard deviation \\ Zero-cross rate\\ $FBE_{1}$\\ $FBE_{2}$\\ $FBE_{3}$\\ $FBE_{6}$\\ $FBE_{10}$\\ $FBE_{12}$\\ $FBE_{13}$\\ Mean of $FBE_{12}$\\ Mean of $FBE_{13}$\end{tabular} & \begin{tabular}[c]{@{}l@{}}$MFCC_{2}$\\ $MFCC_{4}$\\ $MFCC_{7}$\\ $MFCC_{10}$\\ Mean of $MFCC_{7}$\\ $FBE_{2}$\\ $FBE_{9}$\\ Mean of $FBE_{13}$\end{tabular} \\ \hline
\end{tabular}
}
	\end{table}

\subsection{Performance of language-independent and Classifier-independent Features}
Afterwards, in order to select the optimum features subset, the trainings are made by using four subsets of the features: {\textquotedblleft All features\textquotedblright}, {\textquotedblleft 22 best features\textquotedblright}, {\textquotedblleft Special features\textquotedblright} and {\textquotedblleft Common features\textquotedblright} (i.e. language or classifier-independent features subset). The performances of language-independent features using state-of-the-art filter methods for each language is compared with the  performance of language-independent features using individual ranking. This comparison for each language is shown from Table~\ref{T15} to Table~\ref{T17}. Bold values signify that the results obtained using features selected by state-of-the-art filter methods are superated by the individual features ranking of our approach. The lower performance of state-of-the-art filter methods (where feature selection is a pre-processing step) sometimes is possible, because the criterion used for the feature selection is not very well adopted to the classification algorithm. However, assuming the performance obtained from Table~\ref{T15} to Table~\ref{T17},  the optimum feature subset from  Table~\ref{T2} to Table~\ref{T9} in appendix B can be chosen.
On the other hand, from Table~\ref{T18} to Table~\ref{T20}, the performance of classifier-independent features using individual feature ranking is shown for all three feature categories. According to these tables, we can choose the optimum subset from Table~\ref{T10} in appendix B.\\
We implemented another strategy that shows which features are more effective for each emotion. Then in order to analyze which features are related to each emotion, AdaBoost with decision stumps on the Serbian dataset is applied. Since AdaBoost is used to boost the performance of one-level decision trees (stumps) on binary classification problems, we made five datasets per each emotion, with two labels, e.g. happy and not happy. Weighted features per emotions and their recognition rate are shown in Table~\ref{T341}. In this paper, the best subset of features for recognizing each of the emotions is found, and the best for the recognition rate is obtained accordingly. For example, the highest performance in recognizing the happiness emotion, i.e. 90\%, is achieved by using a subset of features which consists of mCC2, FBE9, mCC4, F123-median-mean, pitch and mCC10. The foregoing process is performed for all of the emotions. The best subset of features for recognition of each of the emotions and the corresponding best recognition rate are listed in Table~\ref{T341} . This method can work on the different languages, and it is a useful strategy to check the relation of the features with the particular emotion.

\begin{table*}[ht]
\centering
\caption{The comparison of the performance of language-independent features using state-of-the-art filter methods and our ranking strategy for the Polish dataset}
\label{T15}
\resizebox{0.9\textwidth}{!}{
\begin{tabular}{l|c|c|c|}
\cline{2-4}
\multicolumn{1}{c|}{\textbf{The Polish}} & KNN Performance (\%) & \begin{tabular}[l]{@{}c@{}}M-SVM\\Performance (\%)\end{tabular} & \begin{tabular}[l]{@{}c@{}}Neural Networks\\Performance (\%)\end{tabular} \\ \hline
\multicolumn{1}{|l|}{All features}                        & 53.50                 & 61.00                         & 66.50                \\ \hline
\multicolumn{1}{|l|}{22 best features GR/our approach}                    & 57.00/54.50                 & 58.50/57.50                       & \textbf{55.50}/58.00                  \\ \hline
\multicolumn{1}{|l|}{Common features GR/our approach}                         &  \textbf{44.00}/46.50                 & \textbf{40.00}/51.00                         & 48.00/45.00                  \\ \hline
\multicolumn{1}{|l|}{Special features GR/our approach}                      &   \textbf{47.00}/51.00                   & 59.00/57.50                       & 57.00/52.90                \\ \hline

\multicolumn{1}{|l|}{22 best features IG/our approach}                    & 57.50/54.50                 & \textbf{57.00}/57.50                      & \textbf{53.50}/58.00                  \\ \hline
\multicolumn{1}{|l|}{Common features IG/our approach}                        & 50.50/46.50                 & \textbf{50.50}/51.00                         & 46.00/45.00                  \\ \hline
\multicolumn{1}{|l|}{Special features IG/our approach}                      & 57.00/51.00                   & 59.50/57.50                       & 58.00/52.90                \\ \hline

\multicolumn{1}{|l|}{22 best features RF/our approach}                    & 60.50/54.50                 & 59.50/57.50                       & 62.00/58.00                  \\ \hline
\multicolumn{1}{|l|}{Common features RF/our approach}                        & 50.50/46.50                 & \textbf{47.50}/51.00                         & 45.50/45.00                  \\ \hline
\multicolumn{1}{|l|}{Special features RF/our approach}                      & 58.00/51.00                   & 60.00/57.50                       & 60.50/52.90                \\ \hline

\multicolumn{1}{|l|}{22 best features SU/our approach}                    & \textbf{52.00}/54.50                 & \textbf{55.00}/57.50                       & \textbf{48.00}/58.00                  \\ \hline
\multicolumn{1}{|l|}{Common features SU/our approach}                        &  \textbf{42.50}/46.50                 & \textbf{42.00}/51.00                         & \textbf{42.50}/45.00                  \\ \hline
\multicolumn{1}{|l|}{Special features SU/our approach}                      & \textbf{50.50}/51.00                   & 60.50/57.50                       & 60.50/52.90                \\ \hline

\end{tabular}
}
\end{table*}

\begin{table*}[ht]
\centering
\caption{The comparison of performance of language-independent features using state-of-the-art filter methods and our ranking strategy for the SAVEE dataset}
\label{T16}
\resizebox{0.9\textwidth}{!}{
\begin{tabular}{l|c|c|c|}
\cline{2-4}
\multicolumn{1}{c|}{\textbf{The SAVEE}} & KNN Performance (\%) & \begin{tabular}[l]{@{}c@{}}M-SVM\\Performance (\%)\end{tabular} & \begin{tabular}[l]{@{}c@{}}Neural Networks\\Performance (\%)\end{tabular} \\ \hline
\multicolumn{1}{|l|}{All features}                        & 57.87                 & 60.00                         & 65.55                \\ \hline
\multicolumn{1}{|l|}{22 best features GR/our approach}                    & 61.39/58.61                 & \textbf{51.11}/54.16                       & \textbf{46.39}/53.33                  \\ \hline
\multicolumn{1}{|l|}{Common features GR/our approach}                        & \textbf{42.77}/46.66                 & 43.89/43.61                         &\textbf{43.89}/46.94                  \\ \hline
\multicolumn{1}{|l|}{Special features GR/our approach}                      & \textbf{57.22}/59.16                   & \textbf{48.89}/52.50                       & \textbf{48.05}/48.89                \\ \hline

\multicolumn{1}{|l|}{22 best features IG/our approach}                    & 59.44/58.61                 & \textbf{53.33}/54.16                       & \textbf{47.50}/53.33                  \\ \hline
\multicolumn{1}{|l|}{Common features IG/our approach}                        & 52.22/46.66                 & 45.00/43.61                         & \textbf{45.83}/46.94                  \\ \hline
\multicolumn{1}{|l|}{Special features IG/our approach}                      & \textbf{57.50}/59.16                   & 53.33/52.50                       & \textbf{48.33}/48.89                \\ \hline

\multicolumn{1}{|l|}{22 best features RF/our approach}                    & 60.55/58.61                 & 58.33/54.16                       & 54.16/53.33                  \\ \hline
\multicolumn{1}{|l|}{Common features RF}                        & 57.22/46.66                 & 45.58/43.61                         & \textbf{45.83}/46.94                  \\ \hline
\multicolumn{1}{|l|}{Special features RF/our approach}                      & \textbf{58.05}/59.16                   & \textbf{48.89}/52.50                       & 49.17/48.89                \\ \hline

\multicolumn{1}{|l|}{22 best features SU/our approach}                    & 59.44/58.61                 & 53.33/54.16                       & \textbf{47.22}/53.33                  \\ \hline
\multicolumn{1}{|l|}{Common features SU/our approach}                        & 46.94/46.66                 & 44.16/43.61                         & \textbf{44.44}/46.94                  \\ \hline
\multicolumn{1}{|l|}{Special features SU/our approach}                      & \textbf{55.28}/59.16                   & \textbf{52.22}/52.50                       & \textbf{46.39}/48.89                \\ \hline

\end{tabular}
}
\end{table*}

\begin{table*}[ht]
\centering
\caption{The comparison of performance of language-independent features using state-of-the-art filter methods and our ranking strategy for the Serbian dataset}
\label{T17}
\resizebox{0.9\textwidth}{!}{
\begin{tabular}{l|c|c|c|}
\cline{2-4}
\multicolumn{1}{c|}{\textbf{The Serbian}} & KNN Performance (\%) & \begin{tabular}[l]{@{}c@{}}M-SVM\\Performance (\%)\end{tabular} & \begin{tabular}[l]{@{}c@{}}Neural Networks\\Performance (\%)\end{tabular} \\ \hline
\multicolumn{1}{|l|}{All features}                        & 66.00                 & 70.00                         & 73.30                \\ \hline
\multicolumn{1}{|l|}{22 best features GR/our approach}                    & 74.00/70.67                 & 78.00/70.66                       & 76.00/75.33                  \\ \hline
\multicolumn{1}{|l|}{Common features GR/our approach}                        & 58.67/42.00                 & 52.00/52.00                         & 52.67/52.67                  \\ \hline
\multicolumn{1}{|l|}{Special features GR/our approach}                      & \textbf{64.00}/64.66                   & \textbf{70.67}/76.00                       & 74.00/70.33                \\ \hline

\multicolumn{1}{|l|}{22 best features IG/our approach}                    & 73.33/70.67                 & 76.67/70.66                       & 76.00/75.33                  \\ \hline
\multicolumn{1}{|l|}{Common features IG/our approach}                        & 62.67/42.00                 & 61.33/52.00                         & 61.33/52.67                  \\ \hline
\multicolumn{1}{|l|}{Special features IG/our approach}                      & 64.67/64.66                   & 73.33/\textbf{76.00}                       &74.00/70.33                \\ \hline

\multicolumn{1}{|l|}{22 best features RF/our approach}                    & 70.67/70.67                 & 76.67/70.66                       & \textbf{74.00}/75.33                  \\ \hline
\multicolumn{1}{|l|}{Common features RF/our approach}                        & \textbf{41.33}/42.00                 & \textbf{47.33}/52.00                         & \textbf{46.67}/52.67                  \\ \hline
\multicolumn{1}{|l|}{Special features RF/our approach}                      & 68.00/64.66                   & \textbf{73.33}/76.00                       & 72.67/70.33               \\ \hline

\multicolumn{1}{|l|}{22 best features SU/our approach}                    & 73.33/70.67                 & 76.67/70.66                       & 76.67/75.33                  \\ \hline
\multicolumn{1}{|l|}{Common features SU/our approach}                        & 60.67/42.00                 & \textbf{50.67}/52.00                         & \textbf{50.00}/52.67                  \\ \hline
\multicolumn{1}{|l|}{Special features SU/our approach}                      & \textbf{63.33}/64.66                   & \textbf{71.33}/76.00                       & 70.67/70.33                \\ \hline

\end{tabular}
}
\end{table*}

\begin{table}[ht]
\centering
\caption{Performance of classifier-independent features using our ranking strategy-Polish dataset}
\label{T18}
\resizebox{0.48\textwidth}{!}{
\begin{tabular}{c|c|c|c|c|}
\cline{2-4}
\textbf{The Polish}  & KNN Performance (\%) & \begin{tabular}[l]{@{}c@{}}M-SVM\\Performance (\%)\end{tabular}  & \begin{tabular}[l]{@{}c@{}}Neural Networks\\Performance (\%)\end{tabular} \\ \hline
\multicolumn{1}{|c|}{All features}     & 53.50                 & 61.00                           & 66.50                \\ \hline
\multicolumn{1}{|c|}{22 best features} & 54.50                 & 57.50                         & 58.00                  \\ \hline
\multicolumn{1}{|c|}{Common features}  & 54.50              & 56.50 & 49.00                  \\ \hline
\multicolumn{1}{|c|}{Special features} & 43.00               & 54.50 & 52.50                \\ \hline
\end{tabular}
}
\end{table}
\begin{table}[ht]
\centering
\caption{Performance of classifier-independent features using our ranking strategy for the SAVEE dataset}
\label{T19}
\resizebox{0.48\textwidth}{!}{
\begin{tabular}{c|c|c|c|}
\cline{2-4}
\textbf{The SAVEE}   & KNN Performance (\%) & \begin{tabular}[l]{@{}c@{}}M-SVM\\Performance (\%)\end{tabular}    & \begin{tabular}[l]{@{}c@{}}Neural Networks\\Performance (\%)\end{tabular} \\ \hline
\multicolumn{1}{|c|}{All features}     & 57.87                & 60.00                            & 65.55               \\ \hline
\multicolumn{1}{|c|}{22 best features} & 58.61                & 54.16                         & 53.33               \\ \hline
\multicolumn{1}{|c|}{Common features}     & 58.33                & 52.22 & 48.33               \\ \hline
\multicolumn{1}{|c|}{Special features}   & 63.05                & 51.67 & 49.16               \\ \hline
\end{tabular}
}
\end{table}

\begin{table}[ht]
\centering
\caption{Performance of classifier-independent features using our ranking strategy for the Serbian dataset}
\label{T20}
\resizebox{0.48\textwidth}{!}{
\begin{tabular}{c|c|c|c|}
\cline{2-4}
\textbf{The Serbian} & KNN Performance (\%) & \begin{tabular}[l]{@{}c@{}}M-SVM\\Performance (\%)\end{tabular}   & \begin{tabular}[l]{@{}c@{}}Neural Networks\\Performance (\%)\end{tabular} \\ \hline
\multicolumn{1}{|c|}{All features}     & 66.50                 & 70.00                            & 73.30                \\ \hline
\multicolumn{1}{|c|}{22 best features} & 70.67                & 70.66                         & 75.33               \\ \hline
\multicolumn{1}{|c|}{Common features}     & 66.00                   & 66.00    & 65.33               \\ \hline
\multicolumn{1}{|c|}{Special features}   & 52.00                   & 65.33 & 63.33               \\ \hline
\end{tabular}
}
\end{table}

\begin{table*}[ht]
\centering
\caption{Weighted features per emotions by AdaBoost with decision stump on Serbian dataset}
\label{T341}
\resizebox{0.98\textwidth}{!}{
\begin{tabular}{|c|c|c|c|c|c|}
\hline
\textbf{Emotions:}          & \textbf{Happiness}                                                                                 & \textbf{Angry}                                                                  & \textbf{Fear}                                                                        & \textbf{Sadness}                                                                              & \textbf{Neutral}                                                                                  \\ \hline
\textbf{Weighted features:} & \begin{tabular}[c]{@{}c@{}}mCC2\\ FBE9\\ CC4\\ F123 \_ median \_ mean\\ pitch\\ mCC10\end{tabular} & \begin{tabular}[c]{@{}c@{}}ZCR density\\ FBE2m\\ FBE1\\ F2 \_ median\end{tabular} & \begin{tabular}[c]{@{}c@{}}std\\ ZCR density\\ mCC2\\ CC8\\ FBE10\\ ZCR\end{tabular} & \begin{tabular}[c]{@{}c@{}}FBE2\\ FBE9\\ ZCR density\\ ZCR\\ mFBE13\\ CC2min\end{tabular} & \begin{tabular}[c]{@{}c@{}}CC7\\ FBE13\\ CC8\\ max\\ CC2\\ Intensity\\ mFBE13\\ FBE2\end{tabular} \\ \hline
\textbf{Recognition rate:}  & 90\%                                                                                           & 81.33\%                                                                         & 78.67\%                                                                              & 94.67\%                                                                                   & 91.33\%                                                                                           \\ \hline
\end{tabular}
}
\end{table*}

The classifier-independent feature set chosen based on the three classifiers KNN, MSVM and NN is evaluated using the same classifiers. Moreover, in order to ensure that over-fitting is avoided, they are evaluated using four other classifiers, namely, RF, PCA-KNN, AdaBoost and LogitBoost. They are applied once to all the features, and once more to the mentioned classifiers-independent features, which are listed in Tables \ref{T9}, \ref{T10} and \ref{T11}, for the Polish, SAVEE and Serbian databases, respectively. Tables~\ref{T20:all} and~\ref{T20:common} provide comparisons of the performances of the three previous classifiers and the four new ones using all the features and the classifier-independent ones, respectively.\\
According to Table~\ref{T20:all}, the best three recognition rates using all the features on the Polish, SAVEE and Serbian databases are 68.88\%, 66.50\% and 73\%, which have been achieved by RF, NN and NN classifiers, respectively. The Table also shows that RF can achieve a better performance compared to the previous classifiers, by using all the features on the Polish database. In addition, according to Table~\ref{T20:common}, the proposed method is not prone to over-fitting. More clearly, the selected classifier-independent features have led to improved performances, namely, 69.16\% and 58.50\%, by applying the RF classifier to the Polish and SAVEE databases, respectively. The foregoing values are considerably higher than the performance rates achieved by the classifiers that were the basis of choosing the features.
\begin{table}[ht]
\centering
\caption{Comparison of performances (\%) of new classifiers with previous classifiers based on all features}
\label{T20:all}
\begin{tabular}{c|c|c|c|c|}
\cline{2-4}
\textbf{All Features}                & \begin{tabular}[l]{@{}c@{}}SAVEE\end{tabular}  & \begin{tabular}[l]{@{}c@{}}Polish\end{tabular}   & \begin{tabular}[l]{@{}c@{}}Serbian \end{tabular}  \\ \hline
\multicolumn{1}{|c|}{KNN}            &57.87          &53.50          & 66.50        	\\ \hline   
\multicolumn{1}{|c|}{MSVM}           &60.00          &61.00          & 70.00         	 \\ \hline
\multicolumn{1}{|c|}{NN}             &65.55          &66.50          & 73.30        	 \\ \hline
\multicolumn{1}{|c|}{RF}             &68.88          &63.00          & 55.55        	 \\ \hline
\multicolumn{1}{|c|}{PCA-KNN}        &52.50          &60.55          & 50.00        	 \\ \hline
\multicolumn{1}{|c|}{Adaboost}       &45.55          &36.50          & 43.33        	 \\ \hline
\multicolumn{1}{|c|}{LogitBoost}     &60.27          &60.00          & 50.55         	 \\ \hline
\end{tabular}
\end{table}
\begin{table}[ht]
\centering
\caption{Comparison of performances (\%) of new classifiers with previous classifiers based on common classifier-independent features}
\label{T20:common}
\begin{tabular}{c|c|c|c|c|}
\cline{2-4}
\textbf{Common Features}             & \begin{tabular}[l]{@{}c@{}}SAVEE\end{tabular}  & \begin{tabular}[l]{@{}c@{}}Polish\end{tabular}   & \begin{tabular}[l]{@{}c@{}}Serbian \end{tabular}  \\ \hline
\multicolumn{1}{|c|}{KNN}            &58.33          &54.50          &66.00        	\\ \hline   
\multicolumn{1}{|c|}{MSVM}           &52.22          &56.50          &66.00         	 \\ \hline
\multicolumn{1}{|c|}{NN}             &48.33          &49.00          &65.33        	 \\ \hline
\multicolumn{1}{|c|}{RF}             &69.16          &58.50          &53.33        	 \\ \hline
\multicolumn{1}{|c|}{PCA-KNN}        &52.50          &55.83          &49.44        	 \\ \hline
\multicolumn{1}{|c|}{Adaboost}       &45.83          &36.5           &44.44        	 \\ \hline
\multicolumn{1}{|c|}{LogitBoost}     &57.22          &54.5           &57.22        	 \\ \hline
\end{tabular}
\end{table}
\subsubsection{Comparison of the Language-independent Features with CNN-based Features}\label{S:end}
As aforementioned, in order to assess the efficiency of the proposed set of language-independent features, we compare them with CNN-based features. For this purpose, we combine the samples representing common labels from the Polish, Serbian and SAVEE (English) databases, which are anger, happiness, fear, sadness and neutral, into a unified database.
The audio files are first modified in order to have the same length. Then the spectrogram images are built for all the audio files. Next, each of the images is resized to $\{227\cross{227}\}$ pixels. Then the CNN is applied to the spectrograms, which has resulted in an average recognition rate of 33.63\%.
Afterward, the language-independent features listed in Tables~\ref{T13} and~\ref{T1} are extracted from all the speech signals, and merged into a dataset. Next, the KNN, NN and MSVM classifiers are applied to the dataset, by using 10-fold cross validation for evaluation. As could be seen from Table~\ref{T343}, the language-independent features have led to average recognition rates of 52.36\%, 39\% and 34.71\%, by using the KNN, MSVM and NN classifiers, respectively, which are all higher than the CNN-based average recognition rate. Thus it can be concluded that for a language-independent emotion recognition framework, the proposed combination of paralinguistic acoustic features provides a better distinction between different emotional states, compared to the CNN-based features extracted from spectrograms. 
%
\subsubsection{Language-independent Features -- Polish, SAVEE, Serbian and Italian}
In order to show the flexibility of our approach, after completing the experiment with three corpora (Polish, SAVEE and Serbian), in this experiment the Italian corpora has also been used, and the proposed approach for feature ranking has been applied to it. The Italian corpus is built from voices simulating six emotional states (disgust, fear, anger, joy, surprise and sadness) plus the neutral state. The database includes 84 samples for each emotion, which have been acted by four subjects, i.e. two men and two women.\\
Assuming 22 top features from the ranking list and following the language-independent feature selection strategy, we observed the intersection with features presented in Table~\ref{T13}.
According to Table~\ref{T21}, language-independent features for the Polish, SAVEE, Serbian and Italian corpora are: mean of $MFCC_{1}$, $FBE_{8}$, $FBE_{9}$, Mean of $FBE_{13}$, $FBE_{10}$ and $FBE_{11}$, depending on the method that is used for classification.
\begin{table}[ht]
 	\centering
 	\caption{All selected language-independent features using our ranking strategy--Italian corpora added }
 	\label{T21}
 	\resizebox{0.48\textwidth}{!}{
 	\begin{tabular}{c|l|l|l|}
 \cline{2-4}
 \textbf{Independent features}              & \multicolumn{1}{c|}{\textbf{KNN}}                                       & \multicolumn{1}{c|}{\textbf{MSVM}}                                                                    & \multicolumn{1}{c|}{\textbf{Neural Network}}                                                                                                    \\ \hline
 \multicolumn{1}{|c|}{Language-independent} & \begin{tabular}[c]{@{}l@{}}$MFCC_{1}$\end{tabular} & \begin{tabular}[c]{@{}l@{}} $FBE_{8}$\\ $FBE_{9}$\\ Mean of $FBE_{13}$\end{tabular} & \begin{tabular}[c]{@{}l@{}} $FBE_{9}$\\ $FBE_{10}$\\ $FBE_{11}$\end{tabular} \\ \hline
 \end{tabular}
 }
 \end{table}
We apply the process that was described in Section \ref{S:end} to the Italian database as well, in order to verify that the proposed features perform similarly well on different corpora and different languages. First, we make a new database by combining the samples from the five emotions that are common between all the four available databases. We extract the selected language-independent features that are listed in Table~\ref{T21}. Then we apply each of the KNN, MSVM and NN classifiers. They have resulted in recognition rates of 44.77\% , 40.77\% and 42.22\%, respectively. The foregoing rates show the robustness of the proposed language-independent features against changes of language. Finally, we obtain spectrograms from the wave files in the combined database, as inputs to the CNN. The achieved recognition rate is 32.19\%, which shows that the proposed feature ranking method results in feature sets that are more distinctive than the features selected by powerful deep learning neural networks, i.e. CNNs.
\begin{table*}[t]
	\centering
	\caption{investigation of performance of language-independent features by compering with the CNN results based on combination of all samples of English, Polish and Serbian languages.}
	\label{T343}
	\begin{tabular}{l|c|c|c|c|c|c|}
		\cline{2-7}
		\textbf{}                               & \textbf{Anger} & \textbf{Fear} & \textbf{Happiness} & \textbf{Neutral} & \textbf{Sadness}  & \textbf{Average Recognition Rate}\\ \hline
		\multicolumn{1}{|l|}{\textbf{CNN}}      & 75.82          & 14.29         & 3.30               & 59.34            & 15.38             & 33.63  \\ \hline
		\multicolumn{1}{|l|}{\textbf{KNN}}      & 48.75          & 25.63         & 33.13              & 59.38            & 55.63             & 52.36  \\ \hline
		\multicolumn{1}{|l|}{\textbf{NN}}       & 77.50          & 0.00          & 0.6                & 20.00            & 49.38             & 34.71   \\ \hline
		\multicolumn{1}{|l|}{\textbf{MSVM}}     & 57.5           & 20.0          & 0.00               & 62.50            & 26.25             & 39.12   \\ \hline
		
	\end{tabular}
\end{table*}
%

\section{Conclusion} \label{s4}
In this study, we have proposed a systematic approach for analyzing the state-of-the-art voice quality features to obtain the set of features that can be used for emotion recognition, regardless of the spoken language and method that is adopted for the classification. Adding up more features is always a possibility, and in order to draw a border line, we decided to use only those features which have been employed by other researchers in the field. 
The proposed algorithm is composed of three stages. In the first stage, feature ranking analyzing the state-of-the-art voice quality features is performed. In the second stage, finding the subset of the common features for each language and classifier is described. In the third stage, we compare our approach with state-of-the-art filter methods, comparing their results in terms of recognition rate. According to the obtained results, the optimal sets of features which result in reasonably good performance and are language and classifier-independent could be found.
It is shown that in some cases, since the filter methods are a pre-processing step, the criterion used for the feature selection  is not very well adopted to the classification algorithm, which might result in lower performance. Additionally, in the case of classifier-independent feature selection, using filter methods for feature ranking is not possible, since those methods are independent from the classifier. However, assuming the results obtained using our approach for feature ranking and state-of-the-art filter methods, the optimal sets of features, which result in reasonably good performance and are language and classifier-independent, are found.
On the other hand, wrapper methods, where every subset that is proposed by the subset selection measure is evaluated in the context of the learning algorithm, gives the result that many computationally intensive learning algorithms cannot be used for the classification. Therefore, we didn't use it for the purpose of our research.
Although we have used only three different corpora (English, Polish and Serbian) and three different methods (KNN, M-SVM and Neural Network), the proposed strategy is flexible and can be easily expanded to include an unlimited number of languages and classifiers. It is shown that our method for selecting the most significant language-independent and method-independent features in many cases outperforms state-of-the-art filter methods. At the end, it is shown that classifiers based on the introduced language-independent features outperforms even the CNN method that uses extracted spectrograms from the speech signals.\\
For future work, and in order to obtain a higher emotion recognition accuracy, we will extend our research by using audio-visual data.

\section*{Acknowledgement}
This work has been partially supported by Estonian Research Grant (PUT638), the Estonian Centre of Excellence in IT (EXCITE) funded by the European Regional Development Fund, the Scientific and Technological Research Council of Turkey (T\"UBİTAK) (Proje 1001 - 116E097), the Spanish project TIN2016-74946-P (MINECO/FEDER, UE) and CERCA Programme / Generalitat de Catalunya. We gratefully acknowledge the support of NVIDIA Corporation with the donation of the Titan Xp GPU used for this research.

\bibliographystyle{IEEEtran}
\bibliography{mybibfile}

\onecolumn
\clearpage
\newpage
\section*{Appendix: Features ranking--our approach by using different classifiers and datasets}   \label{AppA}

\begin{table}[H]
\centering
\caption{Numerical label for every feature}
\label{T1}
\resizebox{0.9\textwidth}{!}{
\begin{tabular}{|l|c|c|c|c|c|c|c|}
\cline{2-8}
\hline
{Feature name}     & Label & {Feature name}   & Label & {Feature name}   & Label & {Feature name}   & Label \\
\hline
Max of first formant            & $x_{1}$  & Std              & $x_{22}$ & $MFCC_{11}$             & $x_{43}$ & $FBE_{6}$             & $x_{64}$ \\ 
Max of second formant            & $x_{2}$  & Autocorrolation  & $x_{23}$ & $MFCC_{12}$             & $x_{44}$ & $FBE_{7}$             & $x_{65}$ \\
Max of third formant            & $x_{3}$  & Pitch            & $x_{24}$ & $MFCC_{13}$             & $x_{45}$ & $FBE_{8}$             & $x_{66}$ \\
Min of first formant            & $x_{4}$  & Harm-to-noise   & $x_{25}$ & Mean ($MFCC_{1}$)             & $x_{46}$ & $FBE_{9}$             & $x_{67}$ \\
Min of second formant            & $x_{5}$  & Min              & $x_{26}$ & Mean ($MFCC_{2}$)             & $x_{47}$ & $FBE_{10}$            & $x_{68}$ \\
Min of third formant            & $x_{6}$  & Mean             & $x_{27}$ & Mean ($MFCC_{3}$)             & $x_{48}$ & $FBE_{11}$            & $x_{69}$ \\
Std of first formant            & $x_{7}$  & vari             & $x_{28}$ & Mean ($MFCC_{4}$)             & $x_{49}$ & $FBE_{12}$            & $x_{70}$ \\
\hline
\end{tabular}
}
\end{table}

\begin{table}[H]
	\centering
	\caption{Features ranking--our approach--KNN classifier--Polish dataset}
	\label{T22}
	\resizebox{0.9\textwidth}{!}{
\begin{tabular}{|c|c|c|c|c|c|c|c|}
		\hline
		\multicolumn{1}{|c|}{Feature label} & \multicolumn{1}{c|}{Perf. (\%)} & \multicolumn{1}{c|}{Feature label} & \multicolumn{1}{c|}{Perf. (\%)} & \multicolumn{1}{c|}{Feature label} & \multicolumn{1}{c|}{Perf. (\%)} & \multicolumn{1}{c|}{Feature label} & \multicolumn{1}{c|}{Perf. (\%)} \\ \hline
		$x_{79}$   & 46.00 &    $x_{28}$   & 36.00 &   $x_{52}$  & 25.50 &  $x_{3} $   & 20.00 \\
		$x_{66}$   & 45.50 &    $x_{5} $   & 33.50 &   $x_{37}$  & 25.00 &  $x_{6} $   & 20.00 \\
		$x_{33}$   & 45.00 &    $x_{1} $   & 33.50 &   $x_{54}$  & 24.50 &  $x_{10}$   & 20.00 \\
		$x_{21}$   & 44.50 &    $x_{73}$   & 33.50 &   $x_{36}$  & 24.00 &  $x_{44}$   & 20.00 \\
		$x_{67}$   & 44.00 &    $x_{64}$   & 33.00 &   $x_{47}$  & 24.00 &  $x_{55}$   & 20.00 \\
		$x_{26}$   & 43.50 &    $x_{82}$   & 33.00 &   $x_{18}$  & 23.00 &  $x_{41}$   & 19.50 \\
		$x_{80}$   & 43.00 &    $x_{30}$   & 32.50 &   $x_{35}$  & 22.50 &  $x_{57}$   & 19.50 \\
\hline                             
	\end{tabular}
	}
\end{table}

\begin{table}[H]
	\centering
	\caption{Features ranking--our approach--M-SVM classifier--Polish dataset}
	\label{T23}
	\resizebox{0.9\textwidth}{!}{
\begin{tabular}{|c|c|c|c|c|c|c|c|}
		\hline
		\multicolumn{1}{|c|}{Feature label} & \multicolumn{1}{c|}{Perf. (\%)} & \multicolumn{1}{c|}{Feature label} & \multicolumn{1}{c|}{Perf. (\%)} & \multicolumn{1}{c|}{Feature label} & \multicolumn{1}{c|}{Perf. (\%)} & \multicolumn{1}{c|}{Feature label} & \multicolumn{1}{c|}{Perf. (\%)} \\ \hline
		$x_{71}$   & 50.00 & $x_{30}$  & 36.94 & $x_{79}$  & 31.67 & $x_{36}$  & 26.11 \\
		$x_{21}$   & 48.61 & $x_{46}$  & 36.67 & $x_{72}$  & 31.11 & $x_{37}$  & 25.83 \\
		$x_{29}$   & 48.61 & $x_{19}$  & 36.11 & $x_{14}$  & 30.83 & $x_{73}$  & 25.00 \\
		$x_{70}$   & 48.05 & $x_{69}$  & 35.55 & $x_{43}$  & 30.83 & $x_{17}$  & 24.44 \\
		$x_{26}$   & 45.83 & $x_{78}$  & 35.55 & $x_{44}$  & 30.83 & $x_{57}$  & 24.17 \\
		$x_{22}$   & 44.44 & $x_{66}$  & 35.00 & $x_{55}$  & 30.55 & $x_{1} $  & 23.61 \\
		$x_{59}$   & 43.61 & $x_{76}$  & 35.00 & $x_{24}$  & 30.00 & $x_{6} $  & 23.61 \\
\hline                                
	\end{tabular}
	}
\end{table}

\begin{table}[H]
	\centering
	\caption{Features ranking--our approach--Neural Networks classifier--Polish dataset}
	\label{T24}
	\resizebox{0.9\textwidth}{!}{
\begin{tabular}{|c|c| c|c|c|c|c|c|}
		\hline
		\multicolumn{1}{|c|}{Feature label} & \multicolumn{1}{c|}{Perf. (\%)} & \multicolumn{1}{c|}{Feature label} & \multicolumn{1}{c|}{Perf. (\%)} & \multicolumn{1}{c|}{Feature label} & \multicolumn{1}{c|}{Perf. (\%)} & \multicolumn{1}{c|}{Feature label} & \multicolumn{1}{c|}{Perf. (\%)}\\ \hline
		$x_{	61	}$ &	45.00	& $x_{	81	}	$ &	33.50	& $x_{	34	}	$ &	26.00	& $x_{	16	}$	 & 	20.50	\\ 
		$x_{	46	}$ &	44.50	& $x_{	75	}	$ &	33.00	& $x_{	60	}	$ &	26.00	& $x_{	25	}$	 & 	20.50	\\ 
		$x_{	67	}$ &	44.00	& $x_{	78	}	$ &	33.00	& $x_{	4	}	$ &	25.50	& $x_{	39	}$	 & 	20.50	\\ 
		$x_{	69	}$ &	43.00	& $x_{	65	}	$ &	32.50	& $x_{	6	}	$ &	25.00	& $x_{	43	}$	 & 	20.50	\\ 
		$x_{	68	}$ &	41.50	& $x_{	8	}	$ &	32.00	& $x_{	53	}	$ &	23.50	& $x_{	44	}$	 & 	20.50	\\ 
		$x_{	70	}$ &	41.50	& $x_{	30	}	$ &	32.00	& $x_{	27	}	$ &	23.00	& $x_{	51	}$	 & 	20.50	\\ 
		$x_{	71	}$ &	41.50	& $x_{	31	}	$ &	32.00	& $x_{	35	}	$ &	22.50	& $x_{	1	}$	 & 	20.00	\\
\hline                               
	\end{tabular}
	}
\end{table}

\begin{table}[H]
	\centering
	\caption{Features ranking--our approach--KNN classifier--SAVEE dataset}
	\label{T25}
	\resizebox{0.9\textwidth}{!}{
\begin{tabular}{|c|c| c|c|c|c|c|c|}
		\hline
		\multicolumn{1}{|c|}{Feature label} & \multicolumn{1}{c|}{Perf. (\%)} & \multicolumn{1}{c|}{Feature label} & \multicolumn{1}{c|}{Perf. (\%)} & \multicolumn{1}{c|}{Feature label} & \multicolumn{1}{c|}{Perf. (\%)} & \multicolumn{1}{c|}{Feature label} & \multicolumn{1}{c|}{Perf. (\%)} \\ \hline
		$x_{	71	}$ &	50.00	& $x_{	30	}	$ &	36.94	& $x_{	79	}	$ &	31.67	& $x_{	36	}$	 & 	26.11	\\ 
		$x_{	21	}$ &	48.61	& $x_{	46	}	$ &	36.67	& $x_{	72	}	$ &	31.11	& $x_{	37	}$	 & 	25.83	\\ 
		$x_{	29	}$ &	48.61	& $x_{	19	}	$ &	36.11	& $x_{	14	}	$ &	30.83	& $x_{	73	}$	 & 	25.00	\\ 
		$x_{	70	}$ &	48.05	& $x_{	69	}	$ &	35.55	& $x_{	43	}	$ &	30.83	& $x_{	17	}$	 & 	24.44	\\ 
		$x_{	26	}$ &	45.83	& $x_{	78	}	$ &	35.55	& $x_{	44	}	$ &	30.83	& $x_{	57	}$	 & 	24.17	\\ 
		$x_{	22	}$ &	44.44	& $x_{	66	}	$ &	35.00	& $x_{	55	}	$ &	30.56	& $x_{	1	}$	 & 	23.61	\\ 
		$x_{	59	}$ &	44.44	& $x_{	76	}	$ &	35.00	& $x_{	24	}	$ &	30.00	& $x_{	6	}$	 & 	23.61	\\
\hline                               
	\end{tabular}
	}
\end{table}

\begin{table}[H]
	\centering
	\caption{Features ranking--our approach--M-SVM classifier--SAVEE dataset}
	\label{T26}
	\resizebox{0.9\textwidth}{!}{
\begin{tabular}{|c|c| c|c|c|c|c|c|}
		\hline
		\multicolumn{1}{|c|}{Feature label} & \multicolumn{1}{c|}{Perf. (\%)} & \multicolumn{1}{c|}{Feature label} & \multicolumn{1}{c|}{Perf. (\%)} & \multicolumn{1}{c|}{Feature label} & \multicolumn{1}{c|}{Perf. (\%)} & \multicolumn{1}{c|}{Feature label} & \multicolumn{1}{c|}{Perf. (\%)} \\ \hline
		$x_{	59	}$ &	45.00	& $x_{	65	}	$ &	38.06	& $x_{	4	}	$ &	33.33	& $x_{	49	}$	 & 	33.33	\\ 
		$x_{	22	}$ &	44.16	& $x_{	14	}	$ &	38.05	& $x_{	6	}	$ &	33.33	& $x_{	50	}$	 & 	33.33	\\ 
		$x_{	70	}$ &	43.06	& $x_{	13	}	$ &	37.77	& $x_{	8	}	$ &	33.33	& $x_{	52	}$	 & 	33.33	\\ 
		$x_{	71	}$ &	43.06	& $x_{	15	}	$ &	37.77	& $x_{	9	}	$ &	33.33	& $x_{	53	}$	 & 	33.33	\\ 
		$x_{	61	}$ &	42.50	& $x_{	63	}	$ &	37.50	& $x_{	11	}	$ &	33.33	& $x_{	54	}$	 & 	33.33	\\ 
		$x_{	69	}$ &	42.50	& $x_{	30	}	$ &	36.66	& $x_{	12	}	$ &	33.33	& $x_{	55	}$	 & 	33.33	\\ 
		$x_{	46	}$ &	42.22	& $x_{	74	}	$ &	36.39	& $x_{	18	}	$ &	33.33	& $x_{	57	}$	 & 	33.33	\\
\hline                               
	\end{tabular}
	}
\end{table}

\begin{table}[H]
	\centering
	\caption{Features ranking--our approach--Neural Networks classifier--SAVEE dataset}
	\label{T27}
	\resizebox{0.9\textwidth}{!}{
\begin{tabular}{|c|c|c|c|c|c|c|c|}
		\hline
		\multicolumn{1}{|c|}{Feature label} & \multicolumn{1}{c|}{Perf. (\%)} & \multicolumn{1}{c|}{Feature label} & \multicolumn{1}{c|}{Perf. (\%)} & \multicolumn{1}{c|}{Feature label} & \multicolumn{1}{c|}{Perf. (\%)} & \multicolumn{1}{c|}{Feature label} & \multicolumn{1}{c|}{Perf. (\%)} \\ \hline
		$x_{	59	}$ &	45.28	& $x_{	64	}	$ &	34.17	& $x_{	11	}	$ &	33.33	& $x_{	45	}$	 & 	33.33	\\ 
		$x_{	70	}$ &	44.17	& $x_{	81	}	$ &	34.17	& $x_{	12	}	$ &	33.33	& $x_{	47	}$	 & 	33.33	\\ 
		$x_{	29	}$ &	43.61	& $x_{	28	}	$ &	33.89	& $x_{	17	}	$ &	33.33	& $x_{	48	}$	 & 	33.33	\\ 
		$x_{	22	}$ &	43.33	& $x_{	62	}	$ &	33.89	& $x_{	19	}	$ &	33.33	& $x_{	49	}$	 & 	33.33	\\ 
		$x_{	84	}$ &	43.06	& $x_{	66	}	$ &	33.89	& $x_{	21	}	$ &	33.33	& $x_{	50	}$	 & 	33.33	\\ 
		$x_{	26	}$ &	41.94	& $x_{	82	}	$ &	33.89	& $x_{	23	}	$ &	33.33	& $x_{	51	}$	 & 	33.33	\\ 
		$x_{	83	}$ &	41.67	& $x_{	3	}	$ &	33.61	& $x_{	24	}	$ &	33.33	& $x_{	52	}$	 & 	33.33	\\
\hline                               
	\end{tabular}
	}
\end{table}

\begin{table}[H]
	\centering
	\caption{Features ranking--our approach--KNN classifier--Serbian dataset}
	\label{T28}
	\resizebox{0.9\textwidth}{!}{
\begin{tabular}{|c|c|c|c|c|c|c|c|}
		\hline
		\multicolumn{1}{|c|}{Feature label} & \multicolumn{1}{c|}{Perf. (\%)} & \multicolumn{1}{c|}{Feature label} & \multicolumn{1}{c|}{Perf. (\%)} & \multicolumn{1}{c|}{Feature label} & \multicolumn{1}{c|}{Perf. (\%)} & \multicolumn{1}{c|}{Feature label} & \multicolumn{1}{c|}{Perf. (\%)} \\ \hline
		$x_{	34	}$ &	40.67	& $x_{	42	}	$ &	30.67	& $x_{	80	}	$ &	26.00	& $x_{	81	}$	 & 	20.67	\\ 
		$x_{	47	}$ &	40.67	& $x_{	53	}	$ &	30.67	& $x_{	82	}	$ &	26.00	& $x_{	6	}$	 & 	20.00	\\ 
		$x_{	67	}$ &	40.00	& $x_{	58	}	$ &	30.67	& $x_{	1	}	$ &	25.33	& $x_{	25	}$	 & 	20.00	\\ 
		$x_{	52	}$ &	36.67	& $x_{	69	}	$ &	30.67	& $x_{	71	}	$ &	25.33	& $x_{	51	}$	 & 	20.00	\\ 
		$x_{	60	}$ &	36.67	& $x_{	48	}	$ &	30.00	& $x_{	72	}	$ &	25.33	& $x_{	65	}$	 & 	20.00	\\ 
		$x_{	33	}$ &	36.00	& $x_{	31	}	$ &	29.33	& $x_{	16	}	$ &	24.67	& $x_{	10	}$	 & 	19.33	\\ 
		$x_{	49	}$ &	35.33	& $x_{	38	}	$ &	29.33	& $x_{	30	}	$ &	24.67	& $x_{	15	}$	 & 	19.33	\\
\hline                               
	\end{tabular}
	}
\end{table}

\begin{table}[H]
	\centering
	\caption{Features ranking--our approach--M-SVM classifier--Serbian dataset}
	\label{T29}
	\resizebox{0.9\textwidth}{!}{
\begin{tabular}{|c|c|c|c|c|c|c|c|}
		\hline
		\multicolumn{1}{|c|}{Feature label} & \multicolumn{1}{c|}{Perf. (\%)} & \multicolumn{1}{c|}{Feature label} & \multicolumn{1}{c|}{Perf. (\%)} & \multicolumn{1}{c|}{Feature label} & \multicolumn{1}{c|}{Perf. (\%)} & \multicolumn{1}{c|}{Feature label} & \multicolumn{1}{c|}{Perf. (\%)} \\ \hline
		$x_{	21	}$ &	42.00	& $x_{	83	}	$ &	30.00	& $x_{	57	}	$ &	27.33	& $x_{	31	}$	 & 	23.33	\\ 
		$x_{	52	}$ &	40.00	& $x_{	9	}	$ &	29.33	& $x_{	70	}	$ &	27.33	& $x_{	54	}$	 & 	23.33	\\ 
		$x_{	39	}$ &	38.66	& $x_{	12	}	$ &	29.33	& $x_{	19	}	$ &	26.66	& $x_{	73	}$	 & 	23.33	\\ 
		$x_{	60	}$ &	38.66	& $x_{	15	}	$ &	29.33	& $x_{	1	}	$ &	26.00	& $x_{	13	}$	 & 	22.66	\\ 
		$x_{	68	}$ &	36.66	& $x_{	22	}	$ &	29.33	& $x_{	37	}	$ &	26.00	& $x_{	20	}$	 & 	22.66	\\ 
		$x_{	61	}$ &	36.00	& $x_{	28	}	$ &	29.33	& $x_{	45	}	$ &	26.00	& $x_{	23	}$	 & 	22.66	\\ 
		$x_{	84	}$ &	33.33	& $x_{	33	}	$ &	29.33	& $x_{	50	}	$ &	26.00	& $x_{	48	}$	 & 	22.66	\\
\hline                               
	\end{tabular}
	}
\end{table}

\begin{table}[H]
	\centering
	\caption{Features ranking--our approach--Neural Networks classifier--Serbian dataset}
	\label{T30}
	\resizebox{0.9\textwidth}{!}{
\begin{tabular}{|c|c|c|c|c|c|c|c|}
		\hline
		\multicolumn{1}{|c|}{Feature label} & \multicolumn{1}{c|}{Perf. (\%)} & \multicolumn{1}{c|}{Feature label} & \multicolumn{1}{c|}{Perf. (\%)} & \multicolumn{1}{c|}{Feature label} & \multicolumn{1}{c|}{Perf. (\%)} & \multicolumn{1}{c|}{Feature label} & \multicolumn{1}{c|}{Perf. (\%)}\\ \hline
		$x_{	60	}$ &	42.00	& $x_{	32	}	$ &	29.33	& $x_{	79	}	$ &	27.33	& $x_{	38	}$	 & 	24.00	\\ 
		$x_{	68	}$ &	38.67	& $x_{	78	}	$ &	29.33	& $x_{	82	}	$ &	27.33	& $x_{	44	}$	 & 	24.00	\\ 
		$x_{	36	}$ &	36.67	& $x_{	2	}	$ &	28.67	& $x_{	14	}	$ &	26.67	& $x_{	20	}$	 & 	22.67	\\ 
		$x_{	39	}$ &	35.33	& $x_{	3	}	$ &	28.67	& $x_{	40	}	$ &	26.67	& $x_{	37	}$	 & 	22.67	\\ 
		$x_{	34	}$ &	34.67	& $x_{	12	}	$ &	28.67	& $x_{	57	}	$ &	26.67	& $x_{	76	}$	 & 	22.67	\\ 
		$x_{	69	}$ &	34.67	& $x_{	16	}	$ &	28.67	& $x_{	65	}	$ &	26.67	& $x_{	41	}$	 & 	22.00	\\ 
		$x_{	52	}$ &	34.00	& $x_{	22	}	$ &	28.67	& $x_{	35	}	$ &	26.00	& $x_{	54	}$	 & 	22.00	\\
\hline                               
	\end{tabular}
	}
\end{table}

\begin{table}[H]
	\centering
	\caption{Features ranking--our approach--KNN classifier--Italian dataset}
	\label{T31}
	\resizebox{0.9\textwidth}{!}{
\begin{tabular}{|c|c|c|c|c|c|c|c|}
		\hline
		\multicolumn{1}{|c|}{Feature label} & \multicolumn{1}{c|}{Perf. (\%)} & \multicolumn{1}{c|}{Feature label} & \multicolumn{1}{c|}{Perf. (\%)} & \multicolumn{1}{c|}{Feature label} & \multicolumn{1}{c|}{Perf. (\%)} & \multicolumn{1}{c|}{Feature label} & \multicolumn{1}{c|}{Perf. (\%)}\\ \hline
		$x_{30}$ & 34.01 & $x_{60}$ & 21.59 & $x_{51}$ & 19.22 & $x_{54}$ & 16.84\\
		$x_{31}$ & 32.82 & $x_{64}$ & 21.59 & $x_{76}$ & 19.05 & $x_{16}$ & 16.66\\
		$x_{33}$ & 32.82 & $x_{63}$ & 21.43 & $x_{79}$ & 19.05 & $x_{56}$ & 16.33\\
		$x_{29}$ & 30.27 & $x_{83}$ & 21.08 & $x_{21}$ & 18.70 & $x_{41}$ & 16.16\\
		$x_{28}$ & 26.88 & $x_{77}$ & 20.92 & $x_{45}$ & 18.54 & $x_{44}$ & 15.99\\
		$x_{32}$ & 25.68 & $x_{37}$ & 20.75 & $x_{23}$ & 18.37 & $x_{15}$ & 15.82\\
		$x_{65}$ & 23.98 & $x_{53}$ & 20.75 & $x_{72}$ & 18.37 & $x_{40}$ & 15.82\\
\hline                               
	\end{tabular}
	}
\end{table}

\begin{table}[H]
	\centering
	\caption{Features ranking--our approach--M-SVM classifier--Italian dataset}
	\label{T32}
	\resizebox{0.9\textwidth}{!}{
\begin{tabular}{|c|c|c|c|c|c|c|c|}
		\hline
		\multicolumn{1}{|c|}{Feature label} & \multicolumn{1}{c|}{Perf. (\%)} & \multicolumn{1}{c|}{Feature label} & \multicolumn{1}{c|}{Perf. (\%)} & \multicolumn{1}{c|}{Feature label} & \multicolumn{1}{c|}{Perf. (\%)} & \multicolumn{1}{c|}{Feature label} & \multicolumn{1}{c|}{Perf. (\%)}\\ \hline
		$x_{31}$ & 27.72 & $x_{65}$ & 19.39 & $x_{41}$ & 17.18 & $x_{17}$ & 14.79\\
		$x_{33}$ & 26.70 & $x_{36}$ & 19.22 & $x_{27}$ & 17.01 & $x_{57}$ & 14.79\\
		$x_{30}$ & 26.02 & $x_{74}$ & 19.05 & $x_{45}$ & 16.84 & $x_{72}$ & 14.79\\
		$x_{28}$ & 25.51 & $x_{77}$ & 19.05 & $x_{38}$ & 16.67 & $x_{48}$ & 14.62\\
		$x_{70}$ & 24.66 & $x_{83}$ & 19.04 & $x_{73}$ & 16.67 & $x_{16}$ & 14.45\\
		$x_{71}$ & 24.49 & $x_{62}$ & 18.88 & $x_{12}$ & 16.33 & $x_{24}$ & 14.11\\
		$x_{69}$ & 22.62 & $x_{80}$ & 18.87 & $x_{56}$ & 16.33 & $x_{7}$ & 13.94\\
\hline                               
	\end{tabular}
	}
\end{table}

\begin{table}[H]
	\centering
	\caption{Features ranking--our approach--Neural Networks classifier--Italian dataset}
	\label{T33}
	\resizebox{0.9\textwidth}{!}{
\begin{tabular}{|c|c|c|c|c|c|c|c|}
		\hline
		\multicolumn{1}{|c|}{Feature label} & \multicolumn{1}{c|}{Perf. (\%)} & \multicolumn{1}{c|}{Feature label} & \multicolumn{1}{c|}{Perf. (\%)} & \multicolumn{1}{c|}{Feature label} & \multicolumn{1}{c|}{Perf. (\%)} & \multicolumn{1}{c|}{Feature label} & \multicolumn{1}{c|}{Perf. (\%)}\\ \hline
		$x_{33}$ & 28.57 & $x_{25}$ & 18.53 & $x_{43}$ & 15.98 & $x_{24}$ & 14.11\\
		$x_{30}$ & 26.53 & $x_{80}$ & 18.36 & $x_{54}$ & 15.98 & $x_{35}$ & 14.11\\
		$x_{31}$ & 25.85 & $x_{6}$ & 18.19 & $x_{58}$ & 15.81 & $x_{72}$ & 14.11\\
		$x_{70}$ & 24.83 & $x_{82}$ & 18.19 & $x_{74}$ & 15.81 & $x_{3}$ & 13.95\\
		$x_{28}$ & 24.82 & $x_{8}$ & 17.85 & $x_{49}$ & 15.64 & $x_{16}$ & 13.94\\
		$x_{71}$ & 22.45 & $x_{76}$ & 17.85 & $x_{81}$ & 15.64 & $x_{5}$ & 13.77\\
		$x_{69}$ & 22.27 & $x_{79}$ & 17.85 & $x_{32}$ & 15.47 & $x_{53}$ & 13.77\\
\hline                               
	\end{tabular}
	}
\end{table}

\newpage
\section{Language and classifier-independent features selection } \label{AppB}
\begin{table}[H]
	\centering
	\caption{language-independent features selection--using state-of-the-art filter methods--GR}
	\label{T2}
	\resizebox{0.9\textwidth}{!}{
%
	}
\end{table}

\begin{table}[H]
	\centering
	\caption{language-independent features selection--using state-of-the-art filter methods--IG}
	\label{T3}
	\resizebox{0.9\textwidth}{!}{
%
	}
\end{table}

\begin{table}[H]
	\centering
	\caption{language-independent features selection--using state-of-the-art filter methods--RF}
	\label{T4}
	\resizebox{0.9\textwidth}{!}{
%
	}
\end{table}

\begin{table}[H]
	\centering
	\caption{language-independent features selection--using state-of-the-art filter methods--SU}
	\label{T5}
	\resizebox{0.9\textwidth}{!}{
%
	}
\end{table}

\begin{table}[H]
	\centering
	\caption{language-independent features selection--using our ranking strategy--KNN classifier}
	\label{T6}
	\resizebox{0.9\textwidth}{!}{
%
	}
\end{table}

\begin{table}[H]
	\centering
	\caption{language-independent features selection--using our ranking strategy--M-SVM classifier}
	\label{T7}
	\resizebox{0.9\textwidth}{!}{
%
	}
\end{table}

\begin{table}[H]
	\centering
	\caption{language-independent features selection--using our ranking strategy--Neural Networks classifier}
	\label{T8}
	\resizebox{0.9\textwidth}{!}{
%
	}
\end{table}	

\begin{table}[H]
	\centering
	\caption{classifier-independent features selection--using our ranking strategy--Polish dataset}
	\label{T9}
	\resizebox{0.9\textwidth}{!}{
%
	}
\end{table}

\begin{table}[H]
	\centering
	\caption{classifier-independent features selection--using our ranking strategy--SAVEE dataset}
	\label{T10}
	\resizebox{0.9\textwidth}{!}{
%
	}
\end{table}

\begin{table}[H]
	\centering
	\caption{classifier-independent features selection--using our ranking strategy--Serbian dataset}
	\label{T11}
	\resizebox{0.9\textwidth}{!}{
%
  \\ \hline
		\multicolumn{1}{|l|}{Common features}       & \textbf{$x_{34}$, $x_{36}$, $x_{39}$, $x_{42}$, $x_{52}$, $x_{60}$, $x_{67}$, $x_{84}$}      \\ \hline
	\end{tabular}
	}
\end{table}

\end{document}